\documentclass[aps,prb,amsmath,amssymb,twocolumn,showpacs,superscriptaddress,10pt]{revtex4-2}
\usepackage[english]{babel}
\usepackage{amsmath}
\usepackage{amssymb,amsfonts}
\usepackage{graphicx}
\usepackage[colorlinks=True,linkcolor=blue,citecolor=blue,urlcolor=blue]{hyperref}
\usepackage{bookmark}
\usepackage[dvipsnames]{xcolor}
\usepackage{braket}
\usepackage{nicefrac}
\usepackage{bm}
\usepackage{bbm}
\usepackage{slashed}
\usepackage[normalem]{ulem}
\usepackage{array}
\usepackage{makecell}
\newcolumntype{C}{>{$}c<{$}}
\usepackage{lipsum}
\usepackage{physics}
\usepackage{graphicx}
\usepackage{subfigure}
\usepackage{adjustbox}
\usepackage{bm}
\usepackage{bbold}
\usepackage[dvipsnames]{xcolor}
\usepackage{standalone}
\usepackage{multirow}
\usepackage{tikz}
\usepackage{dsfont}
\usepackage{float}
\usepackage{placeins}
\usepackage{comment}
\usepackage[utf8]{inputenc}
\usepackage[T1]{fontenc}
\usepackage{nicematrix}
\usepackage{leftindex}

\usepackage{siunitx}
\usepackage{physics}

\renewcommand{\Re}{\mathop{\mathrm{Re}}}
\renewcommand{\Im}{\mathop{\mathrm{Im}}}

\DeclareMathAlphabet{\zc}{OT1}{pzc}{m}{it}

  \DeclareUnicodeCharacter{2212}{-}

\usepackage{amsthm}

\newcommand{\ii}{\mathrm{i}}

\newcommand{\ipLR}[2]{\leftindex_\LL {\ip{#1}{#2}} _\RR}
\newcommand{\ipRR}[2]{\leftindex_\RR {\ip{#1}{#2}} _\RR}
\newcommand{\ipRL}[2]{\leftindex_\RR {\ip{#1}{#2}} _\LL}

\newcommand{\melLR}[3]{\leftindex_\LL {\mel{#1}{#2}{#3}} _\RR}
\newcommand{\melRR}[3]{\leftindex_\RR {\mel{#1}{#2}{#3}} _\RR}

\usepackage{booktabs}
\usepackage{multirow}

\newcommand{\LL}{\text{L}}
\newcommand{\RR}{\text{R}}
\newcommand{\FF}{\text{F}}

\usepackage{mathdots}

\begin{document}

\title{Quantum geometrical effects in non-Hermitian systems}

\author{Anton Montag}
\email{anton.montag@mpl.mpg.de}
\affiliation{Advanced Institute for Materials Research (WPI-AIMR), Tohoku University, Sendai 980-8577, Japan}
\affiliation{Max Planck Institute for the Science of Light, Staudtstraße 2, 91058 Erlangen, Germany}
\affiliation{Department of Physics, Friedrich-Alexander Universität Erlangen-Nürnberg, Staudtstraße 7, 91058 Erlangen, Germany}

\author{Tomoki Ozawa}
\email{tomoki.ozawa.d8@tohoku.ac.jp}
\affiliation{Advanced Institute for Materials Research (WPI-AIMR), Tohoku University, Sendai 980-8577, Japan}

\date{\today}

\begin{abstract}
    We explore the relation between quantum geometry in non-Hermitian systems and physically measurable phenomena. We highlight various situations in which the behavior of a non-Hermitian system is best understood in terms of quantum geometry, namely the notion of adiabatic potentials in non-Hermitian systems and the localization of Wannier states in periodic non-Hermitian systems. Further, we show that the non-Hermitian quantum metric appears in the response of the system upon time-periodic modulation, which one can use to experimentally measure the non-Hermitian quantum metric. We validate our results by providing numerical simulations of concrete exemplary systems.
\end{abstract}

\maketitle

\section{Introduction}

Quantum geometry has become a fundamental framework to understand various physical phenomena~\cite{Mead1992,Xiao2010,Dalibard2011,Nakahara2018,Torma2023,Gao2025}.
In Hermitian non-relativistic quantum mechanics, the geometry of quantum states is usually associated with geometric, path dependent phases acquired during the adiabatic evolution along a closed loop in parameter space\cite{Nakahara2018,Simon1983}. 
This geometric phase captures for example the Aharonov-Bohm effect~\cite{Aharnov1959,Wu1975,Berry1984} and the investigation of such geometric properties lead to the discovery of topological states of matter~\cite{Thouless1982,Hasan2010,Qi2011}.
In topological band theory the geometric Berry phase is accumulated by parallel transport of quantum states along closed trajectories in momentum space~\cite{Berry1984,Zak1989}.
The existence of such a geometric phase can be accredited to the existence of a geometrical quantity called the Berry curvature~\cite{Wu1975,Nakahara2018}.
The Berry curvature was first employed in the explanation of the anomalous Hall effect~\cite{Karplus1954,Haldane1988}, and has been used since then in the description of atomic, molecular and optical physics~\cite{Meyer2009,Mironova2013,Jisha2021}, high-energy physics~\cite{Stone2015,Baggio2017} and the theories of polarization and magnetization~\cite{KingSmith1993,Resta1994,Thonhauser2005}.
The Berry curvature is defined as the anti-symmetric part of the complex-valued Hermitian quantum geometric tensor~\cite{Provost1980,Anandan1990,Resta2011,Cayssol2021}.
The remaining symmetric part of the quantum geometric tensors is the quantum metric, which reflects the distance between different quantum states.
Its significance for several physical phenomena recently gained an increasing amount of attention, for example in the theories of flat-band superconductivity~\cite{Peotta2015,Liang2017,Iskin2018}, conductivity in dissipative systems~\cite{Neupert2013,Kolodrubetz2013,Albert2016,Ozawa2018b,Blue2018} and quantum information~\cite{Venuti2007,You2007,Zanardi2007,Dey2012}.

Topological band theory has been expanded to non-Hermitian quantum mechanics, i.e. systems whose states evolve under a Schrödinger equation with a non-Hermitian Hamiltonian~\cite{KatoBook,Bender2007,Ashida2020}. 
While such non-Hermitian frameworks were originally proposed as descriptions of open quantum systems~\cite{Gamow1928,Feshbach1958,Hatano1996,Hatano1997}, they have found many applications in classical systems, such as optics and especially topological photonics~\cite{El2018,Poli2015,Midya2018,Kawabata2019,Ota2020,Bergholtz2021,Okuma2023}.
The Berry phase has been successfully generalized to non-Hermitian quantum mechanics and its implications have been studied both from the theoretical and experimental side~\cite{Garrison1988,Dattoli1990,Liang2013,Shen2018,Singhal2023}.
Currently there is a growing interest in effects arising from the quantum metric in non-Hermitian systems~\cite{Brody2013,Cuerda2024,Chen2024}.
It was shown that the quantum metric appears in semiclassical wavepacket evolution in non-Hermitian systems subject to complex-valued gradient fields~\cite{Silberstein2020,Alon2024,Hu2025} and that it acts as indicator for topological phase transitions in non-Hermitian systems~\cite{Kolodrubetz2013,Gong2018,Chen2024}.

In this paper we investigate several physical phenomena related to the non-Hermitian quantum metric.
These phenomena have in common that their Hermitian counterpart is well described in terms of quantum geometry, as referenced below.
We show how the adiabatic evolution of non-Hermitian systems, separable into a fast and a slow evolving system, can be described by adiabatic potentials~\cite{Berry1984,Lin2009,Lin2009b}, which are determined by the non-Hermitian Berry connection and quantum metric of the fast system.
Further we analyze non-Hermitian Wannier states and prove that their localization is governed by the non-Hermitian quantum metric~\cite{Marzari1997,Vanderbilt2018}.
Finally we present an extraction scheme to measure the quantum metric of non-Hermitian two-level systems. 
We describe a generic protocol which relates the quantum metric to the response to time-dependent perturbations of non-Hermitian systems~\cite{Ozawa2018}. 
For this extraction scheme we derive the non-Hermitian generalization of time-dependent perturbation theory, which to the best of our knowledge has not been done for generic non-hermitian discrete systems.

\section{Formalism and definitions: Quantum geometry for non-Hermitian systems}

We consider generic linear systems whose dynamics are governed by a Schrödinger equation
\begin{equation}\label{eq:Schrodinger}
    \ii \partial_i\ket{\psi} = H_{\bm{\lambda}}\ket{\psi} \, .
\end{equation}
The linear operator $H_{\bm{\lambda}}\neq H_{\bm{\lambda}}^\dagger$ is referred to as non-Hermitian Hamiltonian.
We emphasize that even though we are using terminologies from quantum mechanics, Eq.~(\ref{eq:Schrodinger}) governs various physical systems, not limited to the quantum realm.
Examples of this include classical non-conservative mechanical systems and electric field dynamics inside of complex refractive index materials~\cite{Makris2008,Guo2009,Yokomizo2022}, among others, and the physical interpretation of the wavefunction depends on the observed system.
For an in depth discussion on systems governed by non-Hermitian Hamiltonians we refer to various review articles~\cite{Bender2007,Ashida2020,Bergholtz2021}.
The non-Hermitian Hamiltonian depends on the set of dimensionless parameters $\bm{\lambda}=\{\lambda_j\}_{j=1}^N$ with $N$ being the dimension of the parameter space.
The right eigenstates of the non-Hermitian Hamiltonian $\ket{\psi_{\bm{\lambda}}^n}_\RR$ are found from the eigenvalue equation
\begin{equation}
    H_{\bm{\lambda}}\ket{\psi_{\bm{\lambda}}^n}_\RR = \varepsilon_n(\bm{\lambda})\ket{\psi_{\bm{\lambda}}^n}_\RR .
\end{equation}
The eigenvalues $\varepsilon_n(\bm{\lambda})$ are generally complex-valued functions of the parameter set.
Due to the non-Hermiticity of $H_{\bm{\lambda}}$ the right eigenfunctions associated with different eigenvalues are generally not orthogonal.
Further the adjoint of $\ket{\psi_{\bm{\lambda}}^n}_\RR$ is not a left eigenstate of the non-Hermitian Hamiltonian.
Instead we need to construct a bi-orthogonal basis using the left eigenstates $\leftindex_\LL{\bra{\psi_{\bm{\lambda}}^n}}$~\cite{Brody2013rev}, which are labeled such that 
\begin{equation}
    \leftindex_\LL{\bra{\psi_{\bm{\lambda}}^n}} H_{\bm{\lambda}}=\varepsilon_n(\bm{\lambda})\leftindex_\LL{\bra{\psi_{\bm{\lambda}}^n}} . 
\end{equation}
The right and left eigenstates form the bi-orthogonal basis $\ipLR{\psi_{\bm{\lambda}}^n}{\psi_{\bm{\lambda}}^m}\propto\delta_{nm}$.
We will employ the bi-orthogonal basis and extend the notion of quantum geometry to the non-Hermitian realm.

Recall that for Hermitian systems the quantum geometric tensor is defined as~\cite{Wilczek1989,Gao2025}
\begin{align}
    \chi_{ij,n} &= \Tr\left[P_{\bm{\lambda},n}\left(\partial_i P_{\bm{\lambda},n}\right)\left(\partial_j P_{\bm{\lambda},n}\right)\right] \\
    &=\mel{\partial_i \psi_{\bm{\lambda}}^n}{\mathbb{1}-P_{\bm{\lambda},n}}{\partial_j \psi_{\bm{\lambda}}^n}\, ,
\end{align}
with $\partial_j=\partial/\partial \lambda_j$ and the projector $P_{\bm{\lambda},n}=\ket{\psi_{\bm{\lambda}}^n}\bra{\psi_{\bm{\lambda}}^n}$.
The quantum geometric tensor is Hermitian, and its imaginary anti-symmetric part is the Berry curvature
\begin{equation}
    \Omega_{ij,n} = \ii \left[\chi_{ij,n} -\chi_{ji,n}\right] = \partial_i A_{j,n} - \partial_j A_{i,n},
\end{equation}
where the Berry connection is given by~\cite{Berry1984,Simon1983}
\begin{equation}
    A_{j,n} = \ii \ip{\psi_{\bm{\lambda}}^n}{\partial_j \psi_{\bm{\lambda}}^n}
\end{equation}
with is real symmetric part being the quantum metric~\cite{Provost1980,Anandan1990,Resta1994,Cayssol2021} 
\begin{equation}
    g_{ij,n} = \Re[\chi_{ij,n}] = \frac{1}{2} \left[\chi_{ij,n} +\chi_{ji,n}\right] \, .
\end{equation}
Both Berry curvature and recently also the quantum metric have been related to a plethora of topological transport and response effects in Hermitian condensed matter physics and optics, c.f. Ref.~\cite{Mead1992,Xiao2010,Gao2025}, for a review of the field.

If the quantum geometric tensor is generalized to non-Hermitian Hamiltonians, the choice of eigenstates from the bi-orthogonal basis is important.
The four possible non-Hermitian quantum geometric tensors are defined as
\begin{align}
    \chi_{ij}^{\alpha\beta,n} &= \Tr_\text{bi}\left[P_{\bm{\lambda},n}^{\beta\alpha} \left(\partial_i P_{\bm{\lambda},n}^{\beta\alpha}\right) \left(\partial_j P_{\bm{\lambda},n}^{\beta\alpha}\right) \right] \\
    &= \frac{\leftindex_\alpha {\mel{\partial_i \psi_{\bm{\lambda}}^n}{\mathbb{1}-P_{\bm{\lambda},n}^{\beta\alpha}}{\partial_j \psi_{\bm{\lambda}}^n}}_\beta}{\leftindex_\alpha{\ip{\psi_{\bm{\lambda}}^n}{\psi_{\bm{\lambda}}^n}}_\beta} \, ,
\end{align}
with the bi-orthogonal trace $\Tr_\text{bi}[\cdot] = \sum_n \leftindex_\LL{\mel{\psi_{\bm{\lambda}}^n}{\cdot}{\psi_{\bm{\lambda}}^n}}_\RR$ and the bi-orthogonal projectors
\begin{equation}
    P_{\bm{\lambda},n}^{\beta\alpha} = \frac{\ket{\psi_{\bm{\lambda}}^n}_\beta \leftindex_\alpha{\bra{\psi_{\bm{\lambda}}^n}}}{\leftindex_\alpha{\ip{\psi_{\bm{\lambda}}^n}{\psi_{\bm{\lambda}}^n}}_\beta}\,.
\end{equation}
The non-Hermitian geometric tensor depends only on the eigenstates of a single band and it is invariant under gauge transformations of both left and right eigenvectors.
We note that throughout the paper we do not impose any particular normalization for the left or right eigenvectors.
The main advantage of defining the non-Hermitian quantum geometric tensor without imposing any kind of normalization on the bi-orthogonal basis is that all different choices of left and right eigenvectors can be treated simultaneously.
The reason for this is that generically at most two of the scalar products $\ipLR{\psi_{\bm{\lambda}}^n}{\psi_{\bm{\lambda}}^n}$, $\ipRR{\psi_{\bm{\lambda}}^n}{\psi_{\bm{\lambda}}^n}$ and $\leftindex_\LL{\ip{\psi_{\bm{\lambda}}^n}{\psi_{\bm{\lambda}}^n}}_\LL$ can be normalized at once.
The anti-symmetric parts of the non-Hermitian quantum geometric tensors yields the different non-Hermitian Berry curvatures 
\begin{equation}
    \Omega_{ij,n}^{\alpha\beta} = \ii \left[\chi_{ij,n}^{\alpha\beta}-\chi_{ji,n}^{\alpha\beta}\right] = \partial_i A_{j,n}^{\alpha\beta} - \partial_j A_{i,n}^{\alpha\beta}\, ,
\end{equation}
with the non-Hermitian Berry connection~\cite{Silberstein2020}
\begin{equation}\label{eq:berry_con}
    A_{j,n}^{\alpha\beta} = \ii \frac{\leftindex_\alpha{\ip{\psi_{\bm{\lambda}}^n}{\partial_j \psi_{\bm{\lambda}}^n}}_\beta}{\leftindex_\alpha{\ip{\psi_{\bm{\lambda}}^n}{\psi_{\bm{\lambda}}^n}}_\beta} \, .
\end{equation}
We define the remaining symmetric part of the non-Hermitian quantum geometric tensor as the non-Hermitian quantum metric
\begin{equation}\label{eq:quantum_metric}
    g_{ij,n}^{\alpha\beta} = \frac{1}{2} \left[\chi_{ij,n}^{\alpha\beta} +\chi_{ji,n}^{\alpha\beta}\right] \, .
\end{equation}
We stress that both non-Hermitian Berry curvature and quantum metric are generally complex-valued and that they depend on the four possible combinations of left and right eigenstates.
For the different choices we find general properties given by
\begin{align}
    &g^{\RR\RR}_{ij}\in\mathbb{R}, \quad \; g^{\LL\LL}_{ij} \in\mathbb{R} \\
    &g^{\LL\RR}_{ij} = \left(g^{\RR\LL}_{ij}\right)^*
\end{align}
Further we note that there have been different ways of defining the non-Hermitian quantum metric as the real or even real and symmetric part of the non-Hermitian quantum geometric tensor~\cite{Solnshkov2021,Liao2021,Zhu2021,Zhang2019}. 
We refer to Ref.~\cite{Chen2024} for an excellent discussion of their distinction. 
The definition presented above relates the non-Hermitian quantum geometric tensor to the Berry curvature, independent of the normalization.
It further is the natural choice considering the physical implications of nontrivial Berry curvature and quantum metric in non-Hermitian systems, which will be discussed in the following.

The non-Hermitian Berry curvature has been studied before and has been related to adiabatic growth of eigenstates~\cite{Singhal2023} and the Petermann factor~\cite{Ozawa2025}.
The non-Hermitian quantum geometric tensor is employed to describe semiclassical dynamics of wavepackets in non-Hermitian systems perturbed by potentially complex-valued gradient fields~\cite{Silberstein2020,Hu2025,Alon2024,Behrends2025}. 
Beyond semiclassical dynamics many implications of a non-Hermitian metric have not yet been investigated.
In the following we describe three distinct situations in which non-Hermitian quantum geometry plays a central role.

\section{Adiabatic potentials in non-Hermitian systems}

In Hermitian systems which contain a fast and a slow component, the quantum geometric effects influence the evolution of the wave function~\cite{Berry1984}.
If the fast component adiabatically follows the slowly varying component, the fast evolution can be integrated out resulting in effective scalar and vector potentials, which are given by the quantum metric and Berry connection of the fast degree of freedom.
This has been used in recent years to create and control artificial gauge fields in ultracold atom experiments~\cite{Lin2009,Lin2009b}.
Extending this to the non-Hermitian realm allows us to treat fast-slow systems, which are dissipative and thus have non-Hermitian contributions.

Consider the non-Hermitian Hamiltonian
\begin{equation}
    H_{\bm{x}} = \left(-\frac{\partial_{\bm{x}}^2}{2m} + V(\bm{x}) \right) \mathbb{1}_\FF+ H_\FF(\bm{x})\, ,
\end{equation} 
where $\mathbb{1}_\FF$ is the identity operator on the Hilbert space of the fast degree of freedom and $H_\FF(\bm{x})\neq H_\FF^\dagger(\bm{x})$ is the non-Hermitian Hamiltonian of the fast degree of freedom.
Note that the position $\bm{x}$ enters $H_\FF(\bm{x})$ as a parameter, such that if $H_\FF(\bm{x})$ contains spatial derivatives, they will be not with respect to $\bm{x}$.
The bi-orthogonal eigenstates of $H_\FF(\bm{x})$ are given by the sets $\{\leftindex_\LL{\bra{\phi_n(\bm{x})}}\}$ and $\{\ket{\phi_n(\bm{x})}_\RR\}$ with $\ipLR{\phi_n(\bm{x})}{\phi_m(\bm{x})}=\delta_{nm}$.
Imposing that initially the fast degree of freedom is in a non-degenerate stationary state, meaning in an eigenstate $H_\FF(\bm{x})\ket{\phi_0(\bm{x})}_\RR=\varepsilon_0(\bm{x})\ket{\phi_0(\bm{x})}_\RR$ with $\Im[\varepsilon_0(\bm{x})]=\max_n \Im[\varepsilon_n(\bm{x})]$, we can apply the adiabatic theorem for the evolution of this non-Hermitian system~\cite{Kvitsinsky1991,Nenciu1992}.
Generically there will be regions in the parameter space $\bm{x}\in\mathbb{R}^d$ with different stationary states, but we assume that during the evolution we stay within one such region.
We stress that the results derived here do not hold, if the stationary states change during the evolution.
Further we note that in a generic non-Hermitian system the stationary state is unique, with the notable exception of pseudo-Hermitian systems, where in the unbroken regime all eigenenergies are real thus all states are stationary.
Additionally we impose that the stationary state is non-degenerate everywhere.
Changes in the stationary state and the implication of exceptional steady states are left for discussion in future works.
Within the additional constraint of a non-degenerate stationary state and given the adiabatic evolution of the fast part of the system we can treat both generic and unbroken pseudo-Hermitian systems on the same footing.
The overall state of the system can be written as 
\begin{equation}
    \ket{\psi(\bm{x},t)} = \psi(\bm{x},t) \ket{\phi_0(\bm{x})}_\RR .
\end{equation}
Inserting this state in the Schrödinger equation Eq.~(\ref{eq:Schrodinger}) and applying $\leftindex_\LL{\bra{\phi_0(\bm{x})}}$ from the left, we obtain
\begin{align}
    \ii \partial_t \psi(\bm{x},t) = &\left[-\frac{\partial_{\bm{x}}^2}{2m} + V(\bm{x}) + \varepsilon_0(\bm{x})\right] \psi(\bm{x},t) \nonumber \\
    &-\frac{1}{m}(\partial_{\bm{x}}\psi(\bm{x},t)) \cdot \melLR{\phi_0(\bm{x})}{\partial_{\bm{x}}}{\phi_0(\bm{x})} \\
    &- \frac{1}{2m}\psi(\bm{x},t) \melLR{\phi_0(\bm{x})}{\partial_{\bm{x}}^2}{\phi_0(\bm{x})} .\nonumber 
\end{align}
By introducing the left-right Berry connection of the stationary state as defined in Eq.~(\ref{eq:berry_con}) this equation can be written as
\begin{align}
    \ii \partial_t \psi(\bm{x},t) = &\biggl[-\frac{1}{2m}\left(\partial_{\bm{x}}-\ii \bm{A}_0^{\LL\RR}\right)^2 + V(\bm{x}) + \varepsilon_0(\bm{x})  \\
    &+\frac{\ipLR{\partial_{\bm{x}}\phi_0}{\partial_{\bm{x}}\phi_0} - \bm{A}_0^{\LL\RR} \cdot \bm{A}_0^{\LL\RR} }{2m}\biggr] \psi(\bm{x},t) . \nonumber
\end{align}
Using $\bm{A}_0^{\LL\RR} = (\bm{A}_0^{\RR\LL})^*$ and 
\begin{align}
    (\bm{A}_0^{\RR\LL})^* \cdot \bm{A}_0^{\LL\RR}&=\ipLR{\partial_{\bm{x}}\phi_0}{\phi_0}\cdot\ipLR{\phi_0}{\partial_{\bm{x}}\phi_0}
\end{align}
we find the simpler expression
\begin{align}
    \ii \partial_t \psi(\bm{x},t) = &\biggl[-\frac{1}{2m}\left(\partial_{\bm{x}}-\ii \bm{A}_0^{\LL\RR}\right)^2 + V(\bm{x}) + \label{eq:adia}\\
    &\varepsilon_0(\bm{x}) + \frac{1}{2m}\Tr[g_{0}^{\LL\RR}] \biggr] \psi(\bm{x},t) \, .\nonumber
\end{align}
This shows that the slow wavefunction of a fast-slow non-Hermitian system, in which the fast degree of freedom evolves adiabatically, is governed by a Schrödinger equation where the left-right Berry connection and non-Hermitian quantum metric appear as vector and scalar potential.
\begin{figure*}
    \centering
    \includegraphics[width=\linewidth]{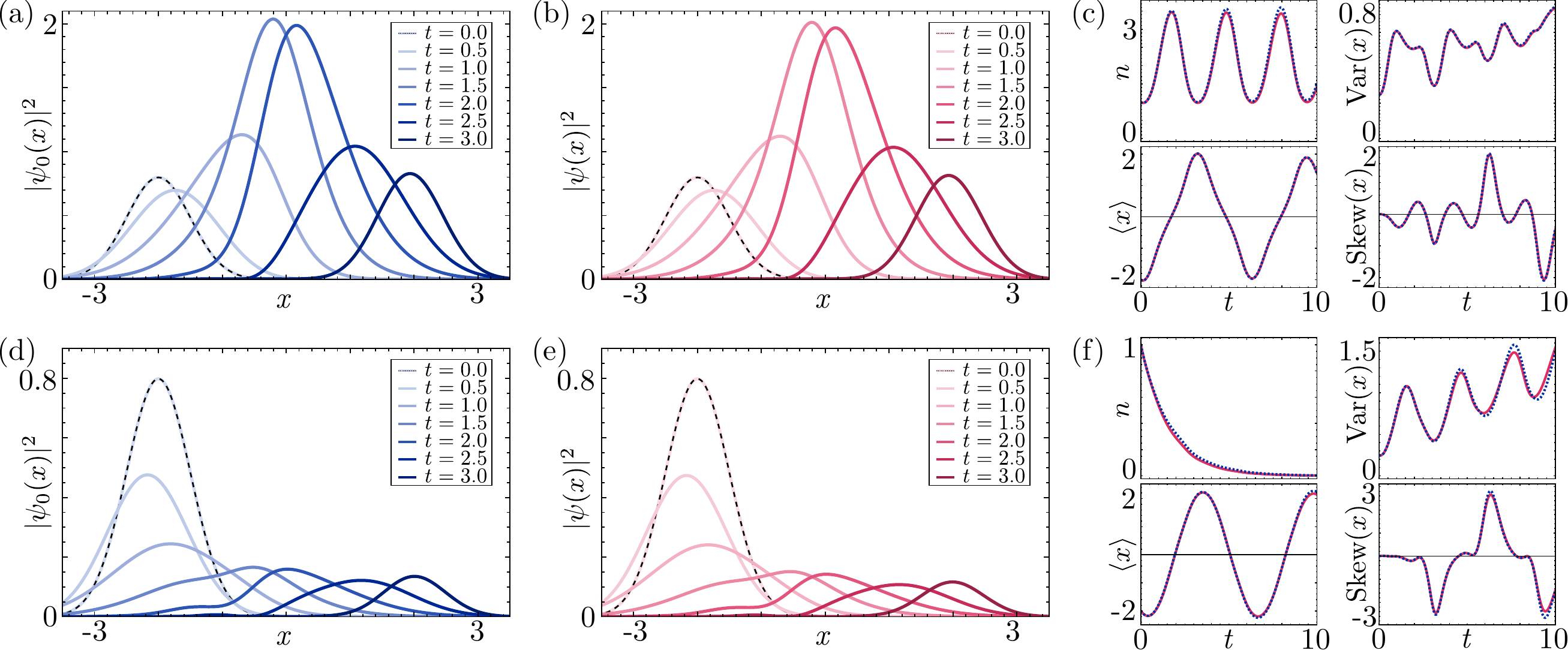}
    \caption{Comparison between the numerical solution of the non-Hermitian two-level system and the solution obtained from the non-Hermitian Schrödinger equation with adiabatic potentials in Eq.~(\ref{eq:adia}). In panels (a) and (d) the evolution of the wavefunction of the two-level systems are shown, where in each case the initial wavefunction (highlighted as black dashed line) is $\psi(x,0)\ket{\phi_0(x)}_\RR$, with $\psi(x,0)=\sqrt[4]{2/\pi} \exp(-(x+2)^2)$ and $\ket{\phi_0(x)}_\RR$ the respective non-decaying eigenstate of $H_\FF(x)$ and $\Tilde{H}_\FF(x)$. Shown in both panels are the projection to the non-decaying state defined by $\psi_0(x,t) = \leftindex_\LL {\ip{\phi_0}{\psi(t)}}$. Panels (b) and (e) show for both systems the solution of $\psi(x,t)$ obtained by considering the adiabatic potentials. Panel (c) compares the norm $n_{(0)}(t)=\int_x |\psi_{(0)}(x,t)|^2$ and the first three central moments of $|\psi_{(0)}(x,t)|^2$ during the evolution governed by $H_\FF(x)$. Here the expectation value is defined at $\langle f(x) \rangle = 1/n(t) \int_x f(x) |\psi_{(0)}(x,t)|^2$. The variance is given by Var$(x)=\langle (x-\langle x\rangle)^2\rangle$ and the skewness by Skew$(x)=\langle (x-\langle x\rangle)^3\rangle/(\text{Var}(x))^{3/2}$. The values obtained from the solution of the two-level system are shown in blue (dashed) and are compared to the values derived from the adiabatic potential solution in red. In panel (f) the same properties are compared for the evolution governed by $\Tilde{H}_\FF(x)$.}
    \label{fig:adia}
\end{figure*}

We show the validity of the non-Hermitian adiabatic potential for a non-Hermitian two level system placed in a one-dimensional harmonic potential $V(x) = x^2/2$, where we set the mass $m=1$.
The fast degree of freedom is governed by
\begin{equation}
    H_\FF(x) = \begin{pmatrix}
        40(1+\ii) & -\frac{80(3+\ii)}{1+x^2} \\
        -80(2-\ii)(1+x^2) &-40(1+3\ii)
    \end{pmatrix} .
\end{equation}
The eigenenergies are given by
\begin{equation}
    \varepsilon_0(x) = -80 \quad \text{and} \quad \varepsilon_1(x) = 80-80\ii \, .
\end{equation}
The system is initialized in its non-decaying state and the evolution is compared to the solution of the non-Hermitian Schrödinger equation in Eq.~(\ref{eq:adia}), with the adiabatic potentials
\begin{align}
    A^{\LL\RR}_0(x) &= -\frac{(1-\ii)x}{1+x^2} \quad \text{and} \\ \Tr[g^{\LL\RR}_0] &= g^{\LL\RR}_{xx,0} = -\frac{2x^2}{(1+x^2)^2} .
\end{align}
In Figure~\hyperref[fig:adia]{1(a)-(c)} the exact evolution and the solution obtained from the adiabatic potentials are compared for an initial Gaussian probability density centered at $x=-2$ with spread $\sigma=0.5$.
The adiabatic assumption holds for the two-level system and the occupation of the decaying state is negligible.
To compare the full solution to the one obtained via the adiabatic potentials, the state $\ket{\psi(t)}$ is projected to the non-decaying state of the non-Hermitian two-level system $\psi_0(t)=\leftindex_\LL {\ip{\phi_0}{\psi(t)}}$.
The shape of the evolving wavefunction and its first three central moments show very good agreement between the numerical solution of the full non-Hermitian two-level system and the solution obtained from the adiabatic potentials.
The state is not decaying but instead the norm shows oscillatory behaviors. 
This periodic gain and loss is a result of the complex non-Hermitian Berry connection, where the sign of $x$ determines whether the system experiences gain or loss.
The quantum metric, on the other hand, is real and therefore does not result in additional decay of the wavefunction.

We consider a different non-Hermitian fast system
\begin{equation}
    \Tilde{H}_\FF(x) = \begin{pmatrix}
        160\ii & -\frac{80(3+\ii)}{1+x^2} \\
        -80(2-\ii)(1+x^2) &-240\ii
    \end{pmatrix} 
\end{equation}
to highlight a qualitatively different behavior of the slow-fast system.
The eigenenergies are identical to the other system
\begin{equation}
    \Tilde{\varepsilon}_0(x) = -80 \quad \text{and} \quad \Tilde{\varepsilon}_1(x) = 80-80\ii \, ,
\end{equation}
however the adiabatic potentials are given by
\begin{align}
    A^{\LL\RR}_0(x) &= -\frac{2x}{1+x^2} \quad \text{and} \\ \Tr[\Tilde{g}^{\LL\RR}_0] &= \Tilde{g}^{\LL\RR}_{xx,0} = -\frac{4(1+\ii)x^2}{(1+x^2)^2},
\end{align}
with complex quantum metric.
We initialize the system again in the non-decaying state with a Gaussian density at $x=-2$ with $\sigma=0.5$.
We expect the imaginary part of the non-Hermitian quantum metric to result in a decay of the wavefunction.
Figure~\hyperref[fig:adia]{1(d)-(f)} confirm again the agreement between solution of the two-level system and the Schrödinger equation with adiabatic potentials.
Further, they show exactly this steady decay of the norm expected from the complex non-Hermitian quantum metric.

The two examples above highlight the flexibility of this approach to design non-Hermitian potentials with desired properties.
A uniquely non-Hermitian phenomenon is the emergence of a complex-valued scalar potential due to the non-Hermitian quantum metric highlighted by the second example.
This results in a decay of the wavepacket constructed from the ground state of the fast system, even though the groundstate has a real eigenenergy everywhere.
By carefully choosing the fast system one can tune between decaying states and oscillating norm of the wavefunction.
This opens a new avenue for shaping wavefunctions using non-Hermitian internal degrees of freedom.

\section{Localization of non-Hermitian Wannier states}

In Hermitian systems the connection between Bloch states and tight-binding descriptions of solid is made by employing Wannier states~\cite{Vanderbilt2018}.
These states are constructed from the Bloch basis and form an orthonormal basis.
The key property of Wannier states is their sharp localization within a single unit cell, thus forming the natural basis for any tight-binding description.
Their localization and spread is in turn governed by the quantum geometry associated with the Bloch states~\cite{Vanderbilt2018,Gao2025}.
In the following we derive similar results for the properties of non-Hermitian Wannier states.

Consider a continuous periodic non-Hermitian systems governed by the non-Hermitian Hamiltonian
\begin{equation}
    H_{\bm{x}} = \frac{1}{2m}\left[-\ii \frac{\partial}{\partial \bm{x}}-\bm{A}(\bm{x})\right]^2 + V(\bm{x}) \, .
\end{equation}
It is defined in terms of the mass $m=m^*$, the vector potential $\bm{A}(\bm{x})\neq\bm{A}^*(\bm{x})$ and the scalar potential $V(\bm{x})\neq V^*(\bm{x})$, resulting in $H_{\bm{x}}\neq H_{\bm{x}}^\dagger$, and the vector and scalar potentials obey the same periodicity
\begin{equation}
    \bm{A}(\bm{x}+\bm{R})=\bm{A}(\bm{x}) \quad \text{and} \quad V(\bm{x}+\bm{R})=V(\bm{x}) ,
\end{equation}
for all real-space lattice vectors $\bm{R}$.
Bloch's theorem hold for non-Hermitian Hamiltonians, therefore the bi-orthogonal eigenstates are found from
\begin{align}
    H_{\bm{x}}\ket{\psi_{\bm{k}}^n(\bm{x})}_\RR &= \varepsilon_n(\bm{k})\ket{\psi_{\bm{k}}^n(\bm{x})}_\RR \quad \text{and} \\
    \leftindex_\LL{\bra{\psi_{\bm{k}}^n(\bm{x})}}H_{\bm{x}} &= \varepsilon_n(\bm{k})\leftindex_\LL{\bra{\psi_{\bm{k}}^n(\bm{x})}} .
\end{align}
Both left and right Bloch eigenstates can be written as
\begin{align}
    \ket{\psi_{\bm{k}}^n(\bm{x})}_\RR &= e^{\ii\bm{k}\cdot\bm{x}}\ket{u_{\bm{k}}^n(\bm{x})}_\RR \quad \text{and} \\
    \leftindex_\LL{\bra{\psi_{\bm{k}}^n(\bm{x})}} &= e^{-\ii\bm{k}\cdot\bm{x}}\leftindex_\LL{\bra{u_{\bm{k}}^n(\bm{x})}} ,
\end{align}
with the plane-wave contribution $e^{\ii\bm{k}\cdot\bm{x}}$ and the periodic part $\ket{u_{\bm{k}}^n(\bm{x})}_\alpha=\ket{u_{\bm{k}}^n(\bm{x}+\bm{R})}_\alpha$ for all lattice vectors $\bm{R}$.
The periodic parts are eigenstates of the Bloch Hamiltonian $H_{\bm{k}}= e^{i\bm{k}\cdot\bm{x}}H_{\bm{x}}e^{-i\bm{k}\cdot\bm{x}}$, which parametrically depends of the quasi-momentum $\bm{k}\in\mathbb{R}^d$. 
Similar to their Hermitian limit, these models are often discussed using a tight-binding approximation.
The connection between continuous model and the tight-binding description can be made by constructing localized bi-orthogonal states, called non-Hermitian Wannier states~\cite{Mochizuki2022}, defined by 
\begin{align}
    \ket{w_{\bm{R}}^n}_\RR &= \int_\text{BZ} \! \frac{d^dk}{(2\pi)^d} \, \frac{e^{-\ii \bm{k}\cdot\bm{R}}\ket{\psi_{\bm{k}}^n}_\RR}{\sqrt{\ipRR{u_{\bm{k}}^n}{u_{\bm{k}}^n}}} \quad \text{and} \\
    \leftindex_\LL{\bra{w_{\bm{R}}^n}} &= \int_\text{BZ} \! \frac{d^dk}{(2\pi)^d} e^{\ii \bm{k}\cdot\bm{R}} \sqrt{\ipRR{u_{\bm{k}}^n}{u_{\bm{k}}^n}} \leftindex_\LL{\bra{\psi_{\bm{k}}^n}} .
\end{align}
They are bi-orthonormal $\ipLR{w_{\bm{R}}^n}{w_{\bm{R}'}^m}=\delta_{nm}\delta_{\bm{R}\bm{R}'}$ and the right Wannier functions of the same band are normalized $\ipRR{w_{\bm{R}}^n}{w_{\bm{R}}^n} = 1$.
The Wannier functions of different lattice vectors $\bm{R}$ are translational images of one another, meaning
for $w_{\bm{R}}^n(\bm{x})=\ip{\bm{x}}{w_{\bm{R}}^n}_\RR$ we find $w_{\bm{R}}^n(\bm{x})=w_{0}^n(\bm{x}-\bm{R})$
A state of the system is usually described in terms of the right eigenstates, thus we are interested in the localization of the right non-Hermitian Wannier states.
First we relate the position matrix elements between right Wannier functions to the right-right Berry connection.
By rewriting
\begin{align}
    &(\hat{\bm{x}}-\bm{R}) \ket{w_{\bm{R}}^n}_\RR \nonumber \\
    = &\int_\text{BZ} \! \frac{d^dk}{(2\pi)^d} \, e^{\ii \bm{k} \cdot (\bm{x}-\bm{R})} \left[\ii \partial_{\bm{k}}\left(\frac{\ket{u_{\bm{k}}^n}_\RR}{\sqrt{\smash[b]{\ipRR{u_{\bm{k}}^n}{u_{\bm{k}}^n}}}}\right)\right]
\end{align}
we find 
\begin{align}
    \melRR{w_0^n}{\hat{\bm{x}}}{w_{\bm{R}}^n} = \int_\text{BZ} \! \frac{d^dk}{(2\pi)^d} e^{-\ii \bm{k}\cdot\bm{R}} \Re\left(\bm{A}_n^{\RR\RR}\right) \, ,\label{eq:Wannier_over}
\end{align}
where we used the definition of the right-right Berry connection Eq.~(\ref{eq:berry_con}).
For the details of the derivation we refer to Appendix~\ref{app1}.
The special case $\bm{R}=0$ simplifies Eq.~(\ref{eq:Wannier_over}) to
\begin{equation}\label{eq:central}
    \melRR{w_0^n}{\hat{\bm{x}}}{w_0^n} = \bm{x}_0 = \int_\text{BZ} \! \frac{d^dk}{(2\pi)^d} \Re\left(\bm{A}_n^{\RR\RR}\right) \, ,
\end{equation}
This integral over the right-right Berry connection determines the center of the Wannier state $\bm{x}_0$ around which it is localized.
We can show, that the average position is gauge-independent.
Non-Hermiticity allows for gauge transformations $\ket{u_{\bm{k}}^n}\rightarrow e^{\ii \phi(\bm{x})} \ket{u_{\bm{k}}^n}$ with a complex-valued $\phi(\bm{x})$. 
The gauge transformation needs to be periodic with respect to the reciprocal lattice vectors $\bm{G}$, thus we find
\begin{equation}
    \phi(\bm{k}+\bm{G})=\phi(\bm{k}) + \bm{G}\cdot\bm{R} \, .
\end{equation}
It is clear that any non-Hermitian gauge transformation can be split into a shift $\phi(\bm{k})=\bm{k}\cdot\bm{R}$, a periodic gauging of the phase $\Re[\phi(\bm{k}+\bm{G})]=\Re[\phi(\bm{k})]$ and a periodic rescaling of the Bloch states $\Im[\phi(\bm{k}+\bm{G})]=\Im[\phi(\bm{k})]$
Crucially the rescaling of the Bloch states, which carries no physical meaning, yields only changes in the imaginary part of $\bm{A}_n^{\RR\RR}$, thus the normalization of the Bloch states does not effect the central position of the Wannier states.
The shift $\phi(\bm{k})=\bm{k}\cdot\bm{R}$ results only in a relabeling of the Wannier states, hence it only changes the assigning of the $\bm{R}=0$ Wannier states.
Lastly the change of the Berry connection due to periodic real part of $\phi(\bm{k})$ vanishes after integrating over the Brillouin zone.
Therefore the center of the Wannier states $\bm{x}_c$ is gauge-invariant.

The spread of the Wannier states is given by 
\begin{align}
    \melRR{w_0^n}{(\Delta\hat{\bm{x}})^2}{w_0^n} = \; &\melRR{w_0^n}{\hat{\bm{x}}^2}{w_0^n}\\
    &-(\melRR{w_0^n}{\hat{\bm{x}}}{w_0^n})^2 \, .\nonumber
\end{align}
The expectation value $\melRR{w_0^n}{\hat{\bm{x}}^2}{w_0^n}$ is given by
\begin{align}
    &\melRR{w_0^n}{\hat{\bm{x}}^2}{w_0^n} \nonumber  \\
    =\; & \int_\text{BZ} \! \frac{d^dk}{(2\pi)^d} \int_\text{BZ} \! \frac{d^dk'}{(2\pi)^d} \frac{\leftindex_\RR{\bra{\psi_{\bm{k}}^n}}}{\sqrt{\smash[b]{\ipRR{u_{\bm{k}}^n}{u_{\bm{k}}^n}}}} \;\hat{\bm{x}}^2 \frac{\leftindex_\RR{\ket{\psi_{\bm{k'}}^n}}}{\sqrt{\smash[b]{\ipRR{u_{\bm{k'}}^n}{u_{\bm{k'}}^n}}}}\nonumber \\
    =\; & \int_\text{BZ} \! \frac{d^dk}{(2\pi)^d} \left(\partial_{\bm{k}}\frac{\leftindex_\RR{\bra{u_{\bm{k}}^n}}}{\sqrt{\smash[b]{\ipRR{u_{\bm{k}}^n}{u_{\bm{k}}^n}}}} \right)\cdot\left(\partial_{\bm{k}}\frac{\ket{u_{\bm{k}}^n}_\RR}{\sqrt{\smash[b]{\ipRR{u_{\bm{k}}^n}{u_{\bm{k}}^n}}}}\right) \nonumber \\
    =\; & \int_\text{BZ} \! \frac{d^dk}{(2\pi)^d} \frac{\ipRR{\partial_{\bm{k}} u_{\bm{k}}^n}{\partial_{\bm{k}} u_{\bm{k}}^n}}{\ipRR{u_{\bm{k}}^n}{u_{\bm{k}}^n}} - \left(\Im[\bm{A}_n^{\RR\RR}]\right)^2 \nonumber \\
    =\; & \int_\text{BZ} \! \frac{d^dk}{(2\pi)^d} \Tr[g_{n}^{\RR\RR}] + \left(\Re[\bm{A}_n^{\RR\RR}]\right)^2 \, .
\end{align}
From this and $\int_\text{BZ} \! \frac{d^dk}{(2\pi)^d} \left(\Re[\bm{A}_n^{\RR\RR}]\right)^2-\left(\int_\text{BZ} \! \frac{d^dk}{(2\pi)^d} \Re[\bm{A}_n^{\RR\RR}]\right)^2\geq 0$ it is clear that 
\begin{equation}
    \melRR{w_0^n}{(\Delta\hat{\bm{x}})^2}{w_0^n} \geq \int_\text{BZ} \! \frac{d^dk}{(2\pi)^d} \Tr[g_{n}^{\RR\RR}] .
\end{equation}
Therefore the spread of non-Hermitian right Wannier states is bounded from below by the right-right non-Hermitian quantum metric integrated over the Brillouin zone.
The right-right non-Hermitian quantum metric is real-valued, resulting in a meaningful bound on the spread of non-Hermitian Wannier states.
We note, that analogous results can be obtained if the left Wannier states are considered instead.
Their center and spread are determined by left-left Berry connection and non-Hermitian quantum metric respectively.

\section{Measurement of non-Hermitian quantum metric through periodic driving}

For Hermitian systems it has been demonstrated that the quantum metric can be extracted from the excitation rate of a quantum system under time-periodic modulation of the parameters.
This result is largely based on time-dependent perturbation theory, which up to our knowledge has not yet been fully generalized to generic non-Hermitian systems.
The perturbation of a Hermitian system with a non-Hermitian time-dependent perturbation has been studied before~\cite{Longhi2017,Pan2020,Geier2022} and high-order perturbative expansions have been used to approximate the short-time dynamics~\cite{Wen2024}.
Additionally, the response of non-Hermitian wavepackets to time-dependent perturbations has recently been derived~\cite{Behrends2025}.
However, transitions between the states of a generic non-Hermitian system induced by time-dependent perturbation have not been described and there is not yet a non-Hermitian analog of Fermi's golden rule~\cite{Landau2013}.
Here, we derive the time-dependent perturbation of a generic non-Hermitian system with a unique steady state.
In Appendix~\ref{app2} we extend this to non-Hermitian systems with multiple non-degenerate stationary states.
We use this method to extract the non-Hermitian right-right quantum metric from  a generic two-band system.

Consider a non-Hermitian system governed by a time-independent Hamiltonian $H_0\neq H_0^\dagger$ and the bi-orthogonal basis sets $\{\leftindex_\LL{\bra{\psi_n}}\}$ and $\{\ket{\psi_n}_\RR\}$ with $H_0\ket{\psi_n}_\RR=(\omega_n-i\gamma_n)\ket{\psi_n}_\RR$ and $\ipLR{\psi_n}{\psi_m}=\delta_{nm}$.
We assume that the system has a unique stationary state $n=0$ with $\gamma_0=\min_n(\gamma_n)$ and $\gamma_0<\gamma_n\;\forall n\neq0$.
The state $\ket{\psi_0}_\RR$ is stationary in the sense that for any generic initial state $\ket{\phi}$ with finite overlap $\leftindex_\LL{\ip{\psi_0}{\phi}}\neq0$ the time evolution of $\ket{\phi}$ will result in a state, in which the contribution of all states $\ket{\psi_n}_\RR$, $n\neq0$ is exponentially suppressed compared to $\ket{\psi_0}_\RR$.
For simplicity we set $\gamma_0=0$.
For systems with multiple nondegenerate stationary states, for example pseudo-Hermitian systems in the unbroken regime, we repeat the derivation in Appendix~\ref{app2}. 
Independent of how the system was initiated at time $t=-\infty$ it will be in the state $\{\ket{\psi_0}_\RR\}$ for finite time $t<0$.
At time $t=0$ a time-dependent non-Hermitian perturbation $\epsilon f(t)H_1$ is turned on, where $H_1$ is a possibly non-Hermitian operator, $\epsilon\ll1$, $f(t)\in\mathbb{R}$.
Our goal is to derive the occupation of all decaying "excited" states $\{\ket{\psi_n}_\RR\}$ to first order in $\gamma$.

We start by expanding the state of the system in right eigenstates of the unperturbed Hamiltonian
\begin{equation}\label{eq:t_dep_state}
    \ket{\psi(t)} = \sum_n c_n(t)e^{-\ii \omega_n t-\gamma_nt} \frac{\ket{\psi_n}_\RR}{\sqrt{\ipRR{\psi_n}{\psi_n}}} \, .
\end{equation}
For $t<0$ the system is in the steady state thus $c_n(t<0)=\delta_{n0}$.
After turning on the time-dependent perturbation the system evolves according to the Schrödinger equation
\begin{equation}
    \ii \partial_t \ket{\psi(t)} = \left(H_0+\epsilon f(t) H_1\right) \ket{\psi(t)} \, .
\end{equation}
By acting with $\leftindex_\LL{\bra{\psi_m}}$ from the left, we find for the complex-valued amplitudes $c_m(t)$ the differential equation
\begin{align}
    \ii\partial_t c_m(t) = \sum_n\Bigl[&\epsilon \sqrt{\tfrac{\ipRR{\psi_m}{\psi_m}}{\ipRR{\psi_n}{\psi_n}}} \melLR{\psi_m}{H_1}{\psi_n} \label{eq:diff_c} \\
    &\times c_n(t) f(t) e^{\ii(\omega_m-\omega_n)t}e^{(\gamma_m-\gamma_n)t}\Bigr] \, . \nonumber
\end{align}
The solution of this equation yields a power series in $\epsilon$, which usually is truncated to achieve the desired precision.
We are interested in the linear response of the system $\propto \epsilon$, thus we insert the unperturbed amplitudes $c_n(t)=\delta_{n0}$ in Eq.~(\ref{eq:diff_c}).
The first order corrections to the amplitudes of the excited states $m\neq0$ are given by
\begin{align}
    c_m(t) = &-\ii\epsilon \sqrt{\tfrac{\ipRR{\psi_m}{\psi_m}}{\ipRR{\psi_0}{\psi_0}}} \melLR{\psi_m}{H_1}{\psi_0} \label{eq:firstorder}\\
    &\times \int_0^t \! dt' f(t') e^{\ii(\omega_m-\omega_0)t'}e^{\gamma_m t'} \nonumber \, .
\end{align}
Note, that while for large $t$ the amplitudes may diverge $\propto e^{\gamma_m t}$, this divergence is exactly canceled, by the decay of the excited states in the non-Hermitian case, cf. Eq~\eqref{eq:t_dep_state}.
Therefore we expect that any periodic time-dependent perturbation will result in a non-zero, periodic occupation of the excited states.
The physical interpretation of this is, that the excitation due to the time-dependent drive and the the decay rate of the excited state are balanced for some non-zero time averaged occupation.
Experimentally this can be accessed by measuring as the time averaged occupation of an excited state given by
\begin{equation}\label{eq:occupation}
    n_m(t) = |P_m^{\RR\LL}\ket{\psi(t)}|^2 = |c_m(t)|^2e^{-2\gamma_m t} \, ,
\end{equation}
where we used a bi-orthogonal projector to extract the occupation of $\ket{\psi_m}_\RR$ in the non-orthogonal right basis.
In the long-time limit for periodic perturbations $f(t)\propto\cos(\omega t)$ the occupation is oscillating around a constant value $\langle n_m(t)\rangle_T$.
This is drastically different to the Hermitian case or to a non-Hermitian system with multiple steady states.
There the occupation grows linearly in time and only the excitation rate can be meaningfully discussed, which results in Fermi's golden rule.
We stress, that the linear response of non-Hermitian system cannot be obtained by using Fermi's golden rule derived for Hermitian systems.
In addition, due to the decay of the excited states, we do not obtain an exact resonance condition here.
Any time-dependent perturbation will result in finite occupation of all excited states.
However, for periodic perturbations $f(t)\propto\cos(\omega t)$ the largest occupation of an excited state is still found for resonant driving $\omega=|\omega_n-\omega_0|$.
For non-Hermitian systems with multiple steady states there will be linear growth of the occupation of a initially empty steady states and this requires exact resonance.

\begin{figure}
    \centering
    \includegraphics[width=\linewidth]{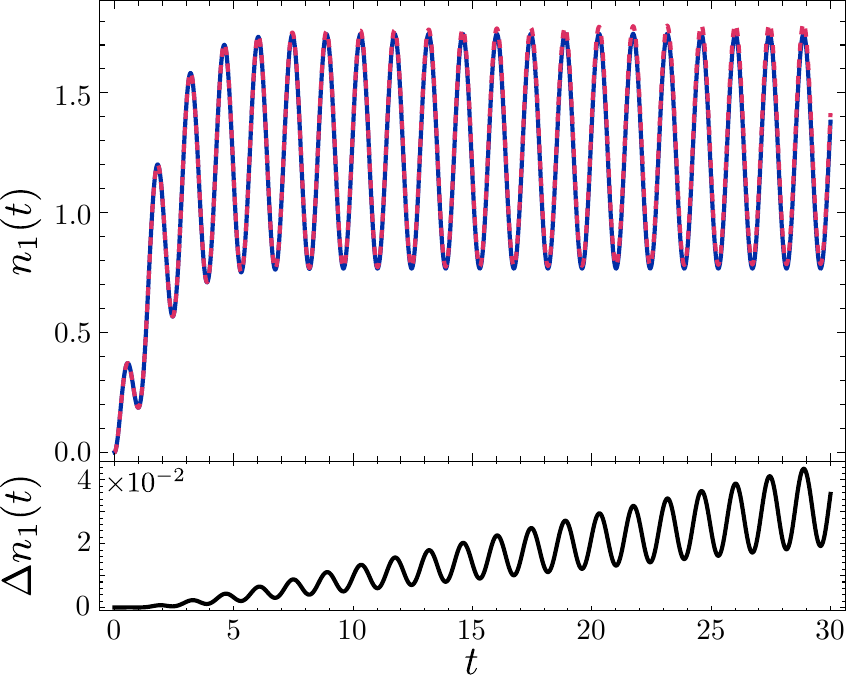}
    \caption{Comparison of occupation of the excited state of the non-Hermitian system defined in Eq.~(\ref{eq:lin_resp_model}) and the time-dependent perturbation given by Eq.~(\ref{eq:perturbation}). The numerical solution $n^\text{num}_1(t)$ (red) is compared with the result obtained by non-Hermitian time-dependent perturbation theory $n^\text{per}_1(t)$ (blue), calculated from Eq.~(\ref{eq:firstorder}). Below the difference $\Delta n_1(t)= n^\text{num}_1(t)-n^\text{per}_1(t)$ is shown.}
    \label{fig:lin_response1}
\end{figure}

To show how accurate the linear response of a non-Hermitian system to the time-dependent perturbation is described by Eq.~(\ref{eq:occupation}), consider the system
\begin{equation}\label{eq:lin_resp_model}
    H_0 = \begin{pmatrix}
        \frac{\ii -1 }{\sqrt{2}} & 1 \\
        1 & - \frac{\ii -1 }{\sqrt{2}}
    \end{pmatrix} + \Im[\sqrt{1-\ii}] \mathbb{1}
\end{equation}
perturbed by $\epsilon f(t) H_1$, with
\begin{equation}\label{eq:perturbation}
    f(t) = 2 \cos(\omega t) \quad \text{and} \quad H_1=\begin{pmatrix}
        0 & -1 \\
        1 & 0
    \end{pmatrix}.
\end{equation}
This model is solved numerically for $\epsilon=0.02$ and $\omega=2.2$ for the initial state being the ground state of $H_0$ given by $\ket{\psi(t=0)}=((1-\ii)/\sqrt{2}+\sqrt{1-\ii},-1)^T$.
In Figure~\ref{fig:lin_response1} the exact time evolution of the occupation of the excited state is compared with the result derived from non-Hermitian time-dependent perturbation theory.
The oscillatory behavior of the occupation is clearly visible.
\begin{figure*}
    \centering
    \includegraphics[width=\textwidth]{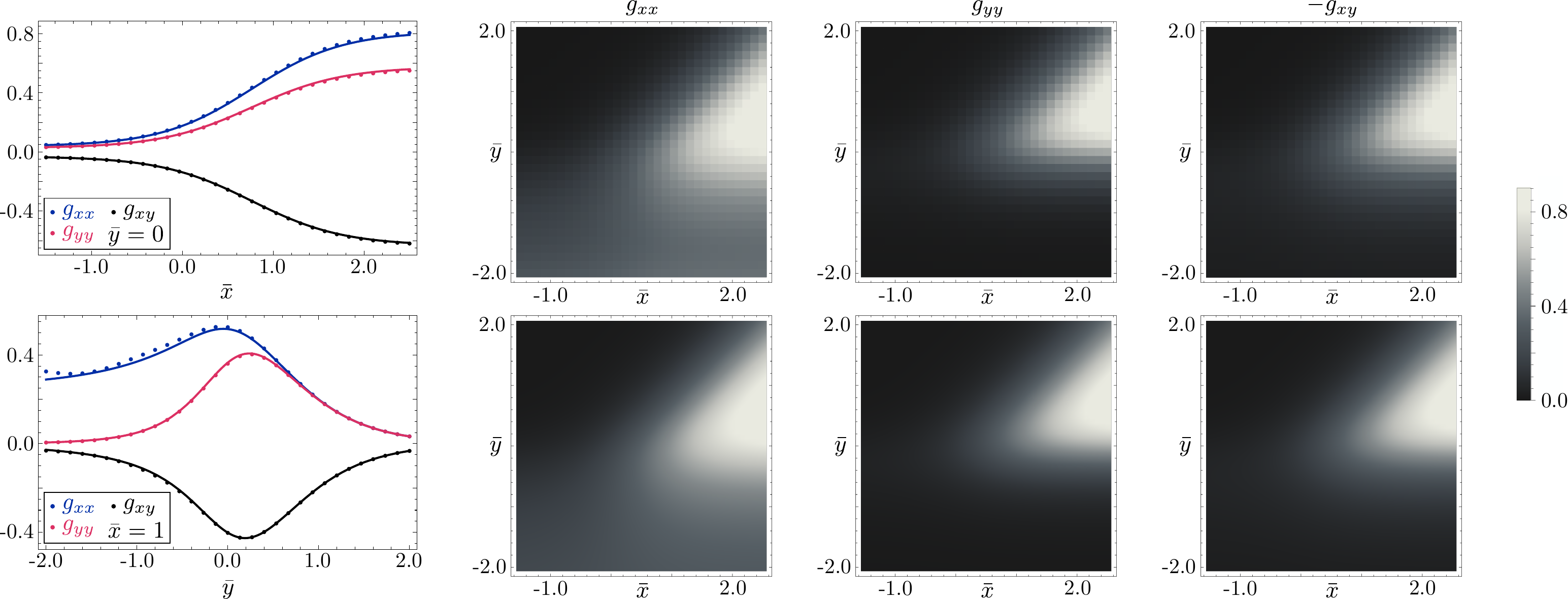}
    \caption{Numerically extracted non-Hermitian quantum metric $g_{ij,0}^{\RR\RR}$ via time-dependent perturbation. The driving frequency is $\omega=2.2$, the perturbation strength is $\epsilon=0.02$ and the time averaging is done over the time interval $t\in[24,24+2\pi/\omega]$. On the left, the measured values of $g_{xx,0}^{\RR\RR}$ (blue dots), $g_{yy,0}^{\RR\RR}$ (red dots), and $g_{xy,0}^{\RR\RR}$ (black dots) are compared with analytically determined exact values (solid lines) for two lines in parameter space. On the right the upper row shows the measured values as density plots and the comparison with the analytical values in the lower plots shows good agreement.}
    \label{fig:quantum_metric}
\end{figure*}

Using the time-dependent perturbation theory developed above, we propose a measurement scheme to extract the right-right non-Hermitian quantum metric.
We consider a generic non-Hermitian 2-band system described by the parameter dependent Hamiltonian $H(\bm{\lambda})$.
Assume that the system has a unique stationary state $\ket{\psi_0}_\RR$.
For simplicity we assume that the $\ket{\psi_0}_\RR$ is not decaying.
Initially the system is prepared in its stationary state at $\bm{\lambda}=\bm{\Bar{\lambda}}$, and to measure the quantum metric at that point in parameter state we modulate one parameter $\lambda_j$ as
\begin{equation}
    \lambda_j(t) = \Bar{\lambda}_j + 2\epsilon \cos(\omega t) \, .
\end{equation}
Given that $\epsilon\ll1$ we Taylor-expand the Hamiltonian
\begin{equation}
    H(\bm{\lambda}(t)) = H(\Bar{\bm{\lambda}}) +2\epsilon \cos(\omega t) \left(\partial_j H(\lambda)\bigr|_{\bm{\lambda}=\Bar{\bm{\lambda}}}\right) \, .
\end{equation}
Using the time-dependent perturbation theory derived before, we find in the long time limit the time-averaged occupation of the excited state given by
\begin{equation}\label{eq:lin_resp_pap}
    \frac{\langle n_1(\omega,t)\rangle_T}{\langle \alpha_1(\omega,t)\rangle_T} = \epsilon^2 \frac{\ipRR{\psi_1}{\psi_1}}{\ipRR{\psi_0}{\psi_0}} \frac{|\melLR{\psi_1}{\partial_j H}{\psi_0}|^2}{|(\omega_1-i\gamma_1)-\omega_0|^2} ,
\end{equation}
where the $\alpha_1(\omega,t)$ depends on the eigenenergies of the unperturbed system and its temporal average is given by 
\begin{align}
    &\frac{1}{\langle \alpha_1(\omega,t)\rangle_T} = \; \frac{\gamma_1^2+\omega^2+(\omega_0-\omega_1)^2}{2((\omega_0-\omega_1)^2+\gamma_1^2)} \\
    &- \frac{2\omega^2(\omega_0-\omega_1)^2}{(\gamma_1^2+\omega^2+(\omega_0-\omega_1)^2)((\omega_0-\omega_1)^2+\gamma_1^2)}\, .\nonumber
\end{align}
To relate the measured occupation to the quantum metric we use $\melLR{\psi_1}{\partial_j H}{\psi_0} = - (\varepsilon_1-\varepsilon_0)\ipLR{\psi_1}{\partial_j \psi_0}$ and the relation
\begin{equation}\label{eq:rewrite_proj}
    \ket{\psi_1}_\LL \leftindex_\LL{\bra{\psi_1}} = \frac{\ipRR{\psi_0}{\psi_0}}{\det(\bm{I})} \left(\mathbb{1}-P_0^{\RR\RR}\right) \, ,
\end{equation}
where we introduced the positive definite Gramian matrix $I_{nm}=\ipRR{\psi_n}{\psi_m}$.
Inserting both relations in Eq.~(\ref{eq:lin_resp_pap}) results in
\begin{equation}
    \frac{\langle n_1(\omega,t)\rangle_T}{\langle \alpha_1(\omega,t)\rangle_T} = \epsilon^2 K_0 g_{jj,0}^{\RR\RR}\, ,
\end{equation}
where we identified the Petermann factor 
\begin{equation}
    K_0=\ipRR{\psi_0}{\psi_0}\leftindex_\LL{\ip{\psi_0}{\psi_0}}_\LL=\frac{\ipRR{\psi_0}{\psi_0}\ipRR{\psi_1}{\psi_1}}{\det(\bm{I})}.
\end{equation}
Note that Eq.~(\ref{eq:rewrite_proj}) is vital for the connection between measurement and quantum metric and that this construction only works for non-Hermitian two-band models.
Even though the proposed schemes are identical our results differ from Ref.~\cite{Chen2024}, where the authors assumed that Fermi's golden rule derived for Hermitian systems holds for non-Hermitian systems in the bi-orthogonal formalism.

Similar to the Hermitian case simultaneous in phase variation of two parameters
\begin{align}
    \lambda_i(t) &= \Bar{\lambda}_i + 2\epsilon \cos(\omega t) \quad \text{and} \\
    \lambda_j(t) &= \Bar{\lambda}_j \pm 2\epsilon \cos(\omega t)
\end{align}
allows us to measure off diagonal elements of the quantum metric by measuring
\begin{equation}
    \frac{\langle n_1^\pm(\omega,t)\rangle_T}{\langle \alpha_1(\omega,t)\rangle_T} = \epsilon^2 K_0 \left[g_{ii,0}^{\RR\RR} \pm 2g_{ij,0}^{\RR\RR}+ g_{jj,0}^{\RR\RR}\right]\, .
\end{equation}
and subsequently taking the difference
\begin{equation}
    \frac{\langle n_1^+(\omega,t)\rangle_T-\langle n_1^-(\omega,t)\rangle_T}{\langle \alpha_1(\omega,t)\rangle_T} = 4\epsilon^2 K_0 g_{ij,0}^{\RR\RR}\, .
\end{equation}

We apply this scheme to the two-band non-Hermitian Hamiltonian 
\begin{equation}
    H(x,y) = \begin{pmatrix}
        \frac{\ii -1}{\sqrt{2}}\, e^y & e^{-x} \\
        e^x & - \frac{\ii -1 }{\sqrt{2}} \, e^y
    \end{pmatrix} + \Im\left[\sqrt{1-\ii\, e^{2y}}\right] \mathbb{1} \, .
\end{equation}
The eigenenergies are given by
\begin{align}
    \varepsilon_0 &= -\Re\left[\sqrt{1-\ii\, e^{2y}}\right] \\
    \varepsilon_1 &= \sqrt{1-\ii\, e^{2y}} + \Im\left[\sqrt{1-\ii\, e^{2y}}\right] .
\end{align}
To test the validity of the proposed scheme numerically, the system is prepared in its ground state $\ket{\psi(0)}=\ket{\psi_0}_\RR/\sqrt{\ipRR{\psi_0}{\psi_0}}$, with
\begin{equation}
    \ket{\psi_0}_\RR = \begin{pmatrix}
        \frac{1-\ii}{\sqrt{2}} \, e^y + \sqrt{1-\ii\, e^{2y}}\\ -e^{x}
    \end{pmatrix} \, .
\end{equation}
The subsequent time evolution of the driven system is simulated numerically while modulating $x(t)=\bar{x} + 2\epsilon\cos{\omega t}$.
The population $n_1(t)$ of the excited decaying state $\ket{\psi_1}_\RR$ is measured after the transients decay and regular oscillation of $n_1(t)$ is observed.
From this the product of non-Hermitian quantum metric element $g_{xx,0}^{\RR\RR}$ and Petermann factor $K_0$ is determined.
Contrary to the Hermitian extraction scheme, where measurements for many frequencies are required~\cite{Ozawa2018}, a single measurement is enough to measure $K_0 g_{xx,0}^{\RR\RR}$ for non-Hermitian systems.
Similarly $y$ and then $x$ and $y$ are modulated to extract both $K_0 g_{yy,0}^{\RR\RR}$ and $K_0 g_{xy,0}^{\RR\RR}$.
In Figure~\ref{fig:quantum_metric} the numerical results are compared with analytical calculations of the non-Hermitian quantum metric as functions of the initial values $\bar{x},\bar{y}$.
All plots show good agreement between the measured non-Hermitian quantum metric and the exact value.
The slight deviations for small $\bar{y}$ originate in the exponentially growing Petermann factor for small $\bar{y}$.

For Hermitian systems this scheme of extracting the quantum metric was introduced for higher-band models~\cite{Ozawa2018}.
However, for non-Hermitian systems it is not possible to extend it to more than two band.
The reason is, that when multiple bands are involved the non-orthogonality prevents us to identify the measured occupation with the right-right non-Hermitian quantum metric.
In both cases we find sums over excited states
\begin{align}
    \text{Hermitian:}& \quad \sum_{n\neq0}^N\ip{\partial_j \psi_0}{\psi_n}\ip{\psi_n}{\partial_j\psi_0} , \label{eq:multiband_herm}\\
    \text{non-Hermitian:}& \quad \sum_{n\neq0}^N \biggl(\frac{\ipRR{\psi_n}{\psi_n}}{\ipRR{\psi_0}{\psi_0}}\ipRL{\partial_j \psi_0}{\psi_n} \label{eq:new_metric} \\
    &\quad\qquad\times\ipLR{\psi_n}{\partial_j\psi_0}\biggr) \nonumber
\end{align}
derived in Appendix~\ref{app3}.
While for Hermitian orthonormal eigenstates
\begin{equation}
    \sum_{n\neq0}^N \op{\psi_n}{\psi_n} = 1-P_0 ,
\end{equation}
we find for a bi-orthogonal basis
\begin{equation}
    \sum_{n\neq0}^N \frac{\ipRR{\psi_n}{\psi_n}}{\ipRR{\psi_0}{\psi_0}} \ket{\psi_n}_\LL \leftindex_\LL{\bra{\psi_n}} \not\propto 1-P^{\RR\RR}_0 \quad \forall N>1 .
\end{equation}
Thus, we cannot relate the measured occupation of excited states to the non-Hermitian quantum metric, even though in the Hermitian limit we recover the quantum metric for multi-band systems.

An open question for future research is whether the quantity similar to Eq.~(\ref{eq:new_metric}) should be included in the list of possible definitions of a non-Hermitian quantum geometric tensor given by
\begin{equation}
    \Tilde{\chi}_{ij,n} = \sum_{m\neq n}\frac{\ipRR{\psi_m}{\psi_m}}{\ipRR{\psi_n}{\psi_n}}\ipRL{\partial_i \psi_n}{\psi_m}\ipLR{\psi_m}{\partial_j\psi_n} \, .
\end{equation}
The Hermitian limit correspond to the proper quantum geometric tensor and it appears in non-Hermitian experiments.
Determining the properties of $\Tilde{\chi}_{ij,n}$ is left to future work.

\section{Conclusions}

We have explored several non-Hermitian systems in which the geometry of quantum states results in measurable physical phenomena.
Our investigation revealed how evolution of fast-slow non-Hermitian systems is governed by an effective Schrödinger equation acting on the state of the slow system, in which the Berry connection and quantum metric of the fast system appear as vector and scalar potential, respectively.
These adiabatic potentials may be non-Hermitian and we showed qualitatively different evolution for real and complex quantum metric, resulting in oscillation or decay of the norm of the wavefunction.
For continuous periodic non-Hermitian systems bi-orthogonal Wannier states may be constructed and we derived the relation between their localization and the quantum geometry associated with the Bloch states, proving that their spread is limited by the quantum metric, similarly to Hermitian Wannier states.
Finally we proposed a scheme to extract the quantum metric from two-level non-Hermitian systems by periodically perturbing the system.
We showed how a time-dependent perturbation excites a non-Hermitian system from its stationary non-decaying state to decaying excited states, derived by considering the linear response of the system.
Contrary to Hermitian systems, in which Fermi's golden rule results in a constant excitation rate but requires precise frequency matching, no exact frequency matching is required in non-Hermitian systems and the decay of the excited states results in a finite occupation of the excited states.
After some initial transient the occupation oscillates around a constant value.
We employed this to determine the quantum metric of a two-level system, by periodically varying the parameters of a two-level system and measuring the time-averaged occupation of the excited state.

Our results highlight the importance of quantum geometry in non-Hermitian settings.
Non-Hermitian quantum geometry has proven before to be an indicator signaling topological phase transitions, but here we showed measurable effects based on quantum geometry, which can be used as a resource to probe and design systems.
Going beyond semiclassical wavepacket dynamics, the notion of non-Hermitian adiabatic potentials enables the usage of artificial gauge fields for non-Hermitian systems.
Adjusting both Berry connection and quantum metric of the system evolving on the fast time-scale, enables us to shape the desired potentials, controlling the evolution of slow system.
This may be used as a non-Hermitian resource for controlling ultracold atomic gasses, in which internal degrees of freedom of the atoms act as fast system.
Furthermore, the proposed experimental extraction scheme of the quantum metric can be applied to topological photonic settings with multiple control parameters.
This makes the quantum metric accessible in a setting, for which non-Hermitian topology has recently found many applications.

\acknowledgments

A.M. would like to thank Flore K. Kunst for providing part of the funding and enabling his research stay at AIMR. A.M. acknowledges funding from both the Max Planck Society Lise Meitner Excellence Program~\mbox{2.0} and the FY2025 JSPS Postdoctoral Fellowships for Research in Japan (Short-term(PE)) with the ID No. PE25273. 
T.O. acknowledges financial support from JSPS KAKENHI Grant No. JP24K00548 and JST PRESTO Grant No. JPMJPR2353.

\appendix

\section{Derivation of Eq.~(\ref{eq:Wannier_over})}
\label{app1}

To calculate the central position of non-Hermitian Wannier states we derive the matrix elements of the position operator $\melRR{w_0^N}{\hat{\bm{x}}}{w_{\bm{R}}^n}$, where we set one lattice index $\bm{R}'=0$ without loss of generality due to the translational properties of the Wannier states.
To evaluate this matrix element we consider
\begin{align}
    & (\hat{\bm{x}}-\bm{R}) \ket{w_{\bm{R}}^n}_\RR \nonumber
    \\ = &\int_\text{BZ} \! \frac{d^dk}{(2\pi)^d} \, \left( -\ii \partial_{\bm{k}}e^{\ii \bm{k} \cdot (\bm{x}-\bm{R})}\right) \frac{\ket{u_{\bm{k}}^n}_\RR}{\sqrt{\smash[b]{\ipRR{u_{\bm{k}}^n}{u_{\bm{k}}^n}}}} \\
    = &\int_\text{BZ} \! \frac{d^dk}{(2\pi)^d} \, e^{\ii \bm{k} \cdot (\bm{x}-\bm{R})} \left[\ii \partial_{\bm{k}}\left(\frac{\ket{u_{\bm{k}}^n}_\RR}{\sqrt{\smash[b]{\ipRR{u_{\bm{k}}^n}{u_{\bm{k}}^n}}}}\right)\right] \\
    = &\int_\text{BZ} \! \frac{d^dk}{(2\pi)^d} \, e^{\ii \bm{k} \cdot (\bm{x}-\bm{R})} \Biggl(\frac{\ii\ket{\partial_{\bm{k}}u_{\bm{k}}^n}_\RR}{\sqrt{\smash[b]{\ipRR{u_{\bm{k}}^n}{u_{\bm{k}}^n}}}} \\
    &\qquad\qquad\qquad-\ii \frac{\ket{u_{\bm{k}}^n}_\RR}{2}\frac{(\partial_{\bm{k}}\ipRR{u_{\bm{k}}^n}{u_{\bm{k}}^n})}{\ipRR{u_{\bm{k}}^n}{u_{\bm{k}}^n}^{3/2}}\Biggr) \nonumber
\end{align}
where we used integration by parts.
The periodicity of $\ket{u_{\bm{k}}^n}_\RR$ cancels all boundary terms appearing in the integration by parts.
By reordering the equation we arrive at
\begin{align}
    \hat{\bm{x}} \ket{w_{\bm{R}}^n}_\RR = &\int_\text{BZ} \! \frac{d^dk}{(2\pi)^d} \, e^{\ii \bm{k} \cdot (\bm{x}-\bm{R})} \Biggl(\bm{R} \frac{\ket{u_{\bm{k}}^n}_\RR}{\sqrt{\smash[b]{\ipRR{u_{\bm{k}}^n}{u_{\bm{k}}^n}}}} \\
    & +\frac{\ii\ket{\partial_{\bm{k}}u_{\bm{k}}^n}_\RR}{\sqrt{\smash[b]{\ipRR{u_{\bm{k}}^n}{u_{\bm{k}}^n}}}}-\ii \frac{\ket{u_{\bm{k}}^n}_\RR}{2}\frac{\partial_{\bm{k}}\ipRR{u_{\bm{k}}^n}{u_{\bm{k}}^n}}{\ipRR{u_{\bm{k}}^n}{u_{\bm{k}}^n}^{3/2}}\Biggr) \nonumber
\end{align}
To obtain the final result we need to make use of of the orthogonality of Bloch functions, which hold for non-Hermitian Bloch functions.
It is derived in Ref.~\cite{Vanderbilt2018} as Eq.~(A.13) and for two generic Bloch functions
\begin{equation}
    \ket{\psi_{\bm{k}}} = e^{\ii \bm{k} \cdot \bm{x}} \ket{u_{\bm{k}}} \quad \text{and} \quad \ket{\chi_{\bm{k}}} = e^{\ii \bm{k} \cdot \bm{x}} \ket{v_{\bm{k}}} , 
\end{equation}
where $\ket{u_{\bm{k}}}$ and $\ket{v_{\bm{k}}}$ are both periodic for all lattice vectors $\bm{R}$, it is given by
\begin{equation}
    \ip{\psi_{\bm{k}'}}{\chi_{\bm{k}}} = (2\pi)^d \delta(\bm{k}-\bm{k}') \ip{u_{\bm{k}'}}{v_{\bm{k}}}.
\end{equation}
We emphasize that this hold for any Bloch state, thus we can apply it the right-right expectation values.
We stress that this result does not apply orthonormality of the non-Hermitian right eigenvectors.
Bloch functions of different bands may still have non-vanishing overlap even though the inner product is diagonal in momentum space.
The position operator matrix elements are given by
\begin{widetext}
\begin{align}
    \melRR{w_0^N}{\hat{\bm{x}}}{w_{\bm{R}}^n} &= \int_\text{BZ} \! \frac{d^dk}{(2\pi)^d}\int_\text{BZ} \! \frac{d^dk'}{(2\pi)^d} \, e^{-\ii \bm{k} \cdot\bm{R}} \Biggl(\bm{R} \frac{\ipRR{\psi_{\bm{k}'}^n}{\psi_{\bm{k}}^n}}{\sqrt{\smash[b]{\ipRR{u_{\bm{k}'}^n}{u_{\bm{k}'}^n}\ipRR{u_{\bm{k}}^n}{u_{\bm{k}}^n}}}}+\frac{\ii\melRR{u_{\bm{k}'}^n}{e^{i(\bm{k}-\bm{k}')\cdot \bm{x}}}{\partial_{\bm{k}}u_{\bm{k}}^n}}{\sqrt{\smash[b]{\ipRR{u_{\bm{k}'}^n}{u_{\bm{k}'}^n}\ipRR{u_{\bm{k}}^n}{u_{\bm{k}}^n}}}} \\
    & \qquad\qquad\qquad\qquad\qquad\qquad\qquad-\ii \frac{\ipRR{\psi_{\bm{k}'}^n}{\psi_{\bm{k}}^n}}{2\sqrt{\smash[b]{\ipRR{u_{\bm{k}'}^n}{u_{\bm{k}'}^n}\ipRR{u_{\bm{k}}^n}{u_{\bm{k}}^n}}}}\frac{\partial_{\bm{k}}\ipRR{u_{\bm{k}}^n}{u_{\bm{k}}^n}}{\ipRR{u_{\bm{k}}^n}{u_{\bm{k}}^n}}\Biggr) \nonumber \\
    &= \int_\text{BZ} \! \frac{d^dk}{(2\pi)^d} e^{-\ii \bm{k} \cdot\bm{R}} \bm{R} + e^{-\ii \bm{k} \cdot\bm{R}}\left(\ii \frac{\ipRR{u_{\bm{k}}^n}{\partial_{\bm{k}} u_{\bm{k}}^n}}{\ipRR{u_{\bm{k}}^n}{u_{\bm{k}}^n}} - \frac{\ii}{2} \frac{\partial_{\bm{k}}\ipRR{u_{\bm{k}}^n}{u_{\bm{k}}^n}}{\ipRR{u_{\bm{k}}^n}{u_{\bm{k}}^n}} \right) \\
    &= \delta_{0,\bm{R}} \bm{R} + \int_\text{BZ} \! \frac{d^dk}{(2\pi)^d} e^{-\ii \bm{k} \cdot\bm{R}} \Re\left(\bm{A}_n^{\RR\RR}\right) . \label{eq:finished}
\end{align}
\end{widetext}
The first term in Eq.~\eqref{eq:finished} never contributes and thus we arrive at the expression found in Eq.~\eqref{eq:Wannier_over} in the main text.
The central position is then found by setting $\bm{R}=0$, which results in
\begin{equation}
    \melRR{w_0^n}{\hat{\bm{x}}}{w_0^n} = \bm{x}_0 = \int_\text{BZ} \! \frac{d^dk}{(2\pi)^d} \Re\left(\bm{A}_n^{\RR\RR}\right) \, ,
\end{equation}
which is given as Eq.~\eqref{eq:central} in the main text.

\section{Time-dependent perturbation theory for non-Hermitian systems with multiple non-degenerate stationary states}
\label{app2}

Here we present a generalization of the time-dependent perturbation theory introduced in the main text covering systems with multiple stationary states.
Prominent examples of such systems are pseudo-Hermitian systems in the unbroken regime. 

Consider a non-Hermitian system governed by a time-independent Hamiltonian $H_0\neq H_0^\dagger$ and the bi-orthogonal basis sets $\{\leftindex_\LL{\bra{\psi_n}}\}$ and $\{\ket{\psi_n}_\RR\}$ with $H_0\ket{\psi_n}_\RR=(\omega_n-i\gamma_n)\ket{\psi_n}_\RR$ and $\ipLR{\psi_n}{\psi_m}=\delta_{nm}$.
For a non-Hermitian system with $N$ states labeled $n\in\{1,\dots,N\}$ we assume that the first $M\leq N$ states are stationary, ie.
\begin{equation}
    \gamma_m = \gamma_0 = \min_n(\gamma_n) \quad \forall 1\leq m\leq M . 
\end{equation}
For simplicity we set $\gamma_0=0$ without loss of generality.
Further we assume that all states are non-degenerate $\omega_n-i\gamma_n\neq\omega_l-i\gamma_l$ for $n\neq l$, because otherwise the system is generally at an exceptional point where perturbation theory may break down.
The state of the system at $t=0$ depends now on the initialization of the system at $t=-\infty$, but still only stationary states can be occupied.
Expanding the initial state in the right-eigenstates of the unperturbed Hamiltonian results in 
\begin{equation}
    \ket{\psi(t)} = \sum_n c_n(t)e^{-\ii \omega_n t-\gamma_nt} \frac{\ket{\psi_n}_\RR}{\sqrt{\ipRR{\psi_n}{\psi_n}}} \, .
\end{equation}
After $t=0$ the system evolves according to the time-dependently perturbed Schrödinger equation
\begin{equation}
    \ii \partial_t \ket{\psi(t)} = \left(H_0+\epsilon f(t) H_1\right) \ket{\psi(t)} \, .
\end{equation}
By acting with $\leftindex_\LL{\bra{\psi_l}}$ from the left, we find for the complex-valued amplitudes $c_l(t)$ the differential equation
\begin{align}
    \ii\partial_t c_l(t) = \sum_n\Bigl[&\epsilon \sqrt{\tfrac{\ipRR{\psi_l}{\psi_l}}{\ipRR{\psi_n}{\psi_n}}} \melLR{\psi_l}{H_1}{\psi_n} \label{eq:diff_csupmat} \\
    &\times c_n(t) f(t) e^{\ii(\omega_l-\omega_n)t}e^{(\gamma_l-\gamma_n)t}\Bigr] \, . \nonumber
\end{align}
The linear response of the system $\propto \epsilon$ is obtained by inserting the unperturbed amplitudes $c_n(t)=c_n(0)$ in Eq.~(\ref{eq:diff_csupmat}) and integrating over time.
For generic initial conditions the resulting occupation of excited and stationary states will have many contributions.
However since the solution is linear in the initial conditions $c_n(0)$ we can focus on the situation where only one of the stationary states is occupied.
We will show that for this initial condition stationary and decaying excited states behave qualitatively differently.

We assume that initially $c_n(0)=\delta_{1n}$.
The amplitudes of the other stationary states $m\neq1$ with $m\leq M$ are given by
\begin{align}
    c_m(t) = &-\ii\epsilon \sqrt{\tfrac{\ipRR{\psi_m}{\psi_m}}{\ipRR{\psi_1}{\psi_1}}} \melLR{\psi_m}{H_1}{\psi_1}  \label{eq:perturb_stationary}\\
    &\times \int_0^t \! dt f(t) e^{\ii(\omega_m-\omega_1)t} \, . \nonumber
\end{align}
We consider now a periodic perturbation $f(t)=\cos(\omega t)$.
The occupation of a stationary state is then given by~\cite{Landau2013}
\begin{align}
    n_m(t) = \;\,&\epsilon^2 \frac{\ipRR{\psi_m}{\psi_m}}{\ipRR{\psi_1}{\psi_1}} |\melLR{\psi_m}{H_1}{\psi_1}|^2 \\
    &\times \left| \int_0^t \! dt \frac{1}{2} e^{\ii(\omega_m-\omega_1-\omega)t}+\frac{1}{2} e^{\ii(\omega_m-\omega_1+\omega)t}\right|^2 \nonumber \\
    \overset{t\rightarrow\infty}{=} &\frac{\pi \epsilon^2t}{2} \frac{\ipRR{\psi_m}{\psi_m}}{\ipRR{\psi_1}{\psi_1}} |\melLR{\psi_m}{H_1}{\psi_1}|^2\\
    &\times\left[\delta(\omega_m-\omega_1-\omega) + \delta(\omega_m-\omega_1+\omega)\right] \,. \nonumber
\end{align}
Thus the occupation of stationary states $m\neq1$ grows linearly in time and we can only meaningfully discuss transition rates 
\begin{align}
    \Gamma_m \equiv &\lim_{t\rightarrow\infty}\frac{n_m(t)}{t} \\
    = &\frac{\pi \epsilon^2}{2} \frac{\ipRR{\psi_m}{\psi_m}}{\ipRR{\psi_1}{\psi_1}} |\melLR{\psi_m}{H_1}{\psi_1}|^2 \\
    &\times \left[\delta(\omega_m-\omega_1-\omega) + \delta(\omega_m-\omega_1+\omega)\right] \, . \nonumber
\end{align}
Note that the transition between stationary states requires matching of the driving frequency to the energy difference of the initial state $n=1$ and the final state $m\neq1$.
This is identical to Fermi's golden rule, however one has to carefully treat the prefactor which depends on the right-eigenstate of the initial state and both the left and the right eigenstate of the final state.
This eigenstate dependence prevents the immediate generalization of the results from Ref.~\cite{Ozawa2018} to extract the quantum metric of unbroken pseudo-Hermitian systems.
Instead similar to the discussion in the main text only the right-right quantum metric can be extracted for two-level systems.

The first order corrections to the amplitudes of the excited states $m>M$ are given by
\begin{align}
    c_m(t) = &-\ii\epsilon \sqrt{\tfrac{\ipRR{\psi_m}{\psi_m}}{\ipRR{\psi_1}{\psi_1}}} \melLR{\psi_m}{H_1}{\psi_1} \\
    &\times\int_0^t \! dt f(t) e^{\ii(\omega_m-\omega_1)t}e^{\gamma_m t} \nonumber \, .
\end{align}
This is identical to the result in the main text and the discussion of this equation will not be repeated here.

For a general initial state all contributions fall in either of the two categories presented above and the full linear response is obtained by summing over all initially occupied states.
The dependence on the matrix element $\melLR{\psi_n}{H_1}{\psi_m}$ contributes linearly in both cases and this might result in unidirectional excitation as discussed for Hermitian $H_0$ and non-Hermitian $H_1$ in Ref.~\cite{Longhi2017}.

\section{Derivation of Eq. (\ref{eq:multiband_herm}) and (\ref{eq:new_metric})}
\label{app3}

First we outline briefly the derivation of Eq.~\eqref{eq:multiband_herm} shown in Ref.~\cite{Ozawa2018}, before generalizing it to the linear response of non-Hermitian higher-band systems with unique steady states given in Eq.~\eqref{eq:new_metric}.

For both cases we consider a parameter-dependent Hamiltonian $H(\bm{\lambda}(t))$, where one parameter is varied as $\lambda_j(t)=\bar{\lambda}_j+2\epsilon\cos(\omega t)$, and we assume that the initial state of the system at $t=0$ is a non-degenerate (right) eigenstate $\ket{\psi_0}$ of $H(\bar{\bm{\lambda}})$.
In a Hermitian system the excitation rate $\Gamma_n(\omega)$ of the system from the initial state to the final state $\ket{\psi}_n$ is given by Fermi's golden rule resulting in
\begin{align}
    \Gamma_f(\omega) = &2\pi\epsilon^2\left|\frac{\mel{\psi_n}{\partial_j H(\bar{\bm{\lambda}})}{\psi_0}}{\omega}\right|^2 \\
    &\times\left(\delta(\varepsilon_n-\varepsilon_0-\omega) + \delta(\varepsilon_n-\varepsilon_0+\omega)\right) \, . \nonumber
\end{align}
What is then measured to determine the quantum metric is the integrated total excitation rate 
\begin{equation}
    \Gamma_\text{int} = \sum_n\int d\omega \; \Gamma_f(\omega) = 2\pi \epsilon^2 \sum_n \frac{|\mel{\psi_n}{\partial_j H(\bar{\bm{\lambda}})}{\psi_0}|^2}{(\varepsilon_n - \varepsilon_0)^2} \, .
\end{equation}
Subsequently employing the identity $\mel{\psi_n}{\partial_j H(\bar{\bm{\lambda}})}{\psi_0} = -(\varepsilon_n - \varepsilon_0)\ip{\psi_n}{\partial_j \psi_0}$ one finds
\begin{equation}
    \Gamma_\text{int} = 2\pi\epsilon^2\sum_n\ip{\partial_j \psi_0}{\psi_n}\ip{\psi_n}{\partial_j \psi_0} \, ,
\end{equation}
which is the relation referenced in Eq~\eqref{eq:multiband_herm}.
From this the quantum metric can be determined.

For non-Hermitian higher-band system it is straight forward to generalize Eq.~\eqref{eq:lin_resp_pap} to multiple decaying excited states $\ket{\psi_m}_\RR$ 
\begin{equation}
    \frac{\langle n_m(\omega,t)\rangle_T}{\langle \alpha_m(\omega,t)\rangle_T} = \epsilon^2 \frac{\ipRR{\psi_m}{\psi_m}}{\ipRR{\psi_0}{\psi_0}} \frac{|\melLR{\psi_m}{\partial_j H}{\psi_0}|^2}{|(\omega_m-i\gamma_m)-\omega_0|^2} 
\end{equation}
giving the time-averaged occupations.
Analogous to the two-band model we use $\melLR{\psi_m}{\partial_j H}{\psi_0} = - (\varepsilon_m-\varepsilon_0)\ipLR{\psi_m}{\partial_j \psi_0}$ resulting in
\begin{equation}
    \frac{\langle n_m(\omega,t)\rangle_T}{\langle \alpha_m(\omega,t)\rangle_T} = \epsilon^2 \frac{\ipRR{\psi_m}{\psi_m}}{\ipRR{\psi_0}{\psi_0}} \ipRL{\partial_j \psi_0}{\psi_m}\ipLR{\psi_m}{\partial_j \psi_0} 
\end{equation}
Similarly to the Hermitian case we might now try to find the non-Hermitian quantum metric by summing over all excited states
\begin{align}
    \sum_m \frac{\langle n_m(\omega,t)\rangle_T}{\langle \alpha_m(\omega,t)\rangle_T} = \epsilon^2\sum_m \biggl(&\frac{\ipRR{\psi_m}{\psi_m}}{\ipRR{\psi_0}{\psi_0}} \ipRL{\partial_j \psi_0}{\psi_m} \\
    &\times\ipLR{\psi_m}{\partial_j \psi_0} \biggr)\nonumber \, ,
\end{align}
resulting in Eq.~\eqref{eq:new_metric} and the discussion following it in the main text.

\bibliography{references.bib}

\end{document}